\documentclass[12pt]{iopart}
\usepackage{graphicx}
\usepackage{dcolumn}
\usepackage{bm}
\usepackage{graphics}
\usepackage{subfigure}
\usepackage{graphicx}
\usepackage{epsfig}
\usepackage{epstopdf}
\usepackage{xcolor}
\usepackage{amssymb}
\expandafter\let\csname equation*\endcsname\relax
\expandafter\let\csname endequation*\endcsname\relax
\usepackage{amsmath}

\usepackage{etoolbox}

\makeatletter
\newcommand{\mainmatter}{%
  \setcounter{footnote}{0}%
  \patchcmd{\@makefntext}{\fnsymbol}{\arabic}{}{}%
  \patchcmd{\@thefnmark}{\fnsymbol}{\arabic}{}{}%
  \def\@makefnmark{\textsuperscript{\arabic{footnote}}}%
}
\makeatother

\begin{document}

\title[Hydrodynamics of a superfluid smectic]{Hydrodynamics of a superfluid smectic}
\author{Johannes Hofmann$^{1}$*, Wilhelm Zwerger$^{2}$*}

\address{$^1$ Department of Physics, Gothenburg University, 41296 Gothenburg, Sweden}
\address{$^2$ Technische Universit\"at M\"unchen, Physik Department, James-Franck-Strasse, 85748 Garching, Germany}

\ead{johannes.hofmann@physics.gu.se, zwerger@tum.de}

\begin{abstract}
We determine the hydrodynamic modes of the superfluid analog of a smectic-A phase in liquid crystals, i.e., a state in which both gauge invariance and translational invariance along a single direction are spontaneously broken. Such a superfluid smectic provides an idealized description of the incommensurate supersolid state realized in Bose-Einstein condensates with strong dipolar interactions as well as of the stripe phase in Bose gases with spin-orbit coupling. We show that the presence of a finite normal fluid density in the ground state of these systems gives rise to a well-defined second-sound type mode even at zero temperature. It replaces the diffusive permeation mode of a normal smectic phase and is directly connected with the classic description of supersolids by Andreev and Lifshitz in terms of a propagating defect mode. An analytic expression is derived for the two sound velocities that appear in the longitudinal excitation spectrum. It only depends on the low-energy parameters associated with the two independent broken symmetries, which are the effective layer compression modulus and the superfluid fraction. 
\end{abstract}

\maketitle

\mainmatter

\section{Introduction}

The question whether superfluidity might persist even in a solid state has a long history. It was briefly discussed
in the classic paper on off-diagonal long range order by Penrose and Onsager~\cite{penrose56}, concluding that no 
supersolid phase is possible because mobile defects or interstitials would always be frozen out at zero temperature. The
idea was taken up by Andreev and Lifshitz~\cite{andreev69} who showed that supersolids are at least a theoretical possibility 
and are characterized by a sound-like rather than diffusive propagation of defects or interstitials. An upper bound on the associated 
superfluid fraction $f_s<1$ that only involves the inhomogeneous density profile was given by Leggett~\cite{leggett70}. It shows 
that in ground states with broken translation invariance superfluidity is favored by a small value of the density contrast. The work by Leggett has been extended by Prokof'ev and Svistunov~\cite{prokofev05} who showed that the existence of zero-point 
vacancies or interstitials is indeed a necessary condition for supersolids. As a consequence, supersolid states in a commensurate situation, i.e., with an integer number of atoms per unit cell, require a fine-tuning to a vanishing value of 
the defect density and are thus not generic. Experimentally, a supersolid phase with a tiny superfluid fraction $f_s \lesssim 10^{-4}$ was inferred from a reduced value of the 
rotational inertia in $^4$He below $250\, {\rm mK}$ by Kim and Chan~\cite{kim04a,kim04b}. On the basis of a number of further 
experiments~\cite{rittner06} and microscopic ab-initio calculations~\cite{clark06,boninsegni06}, however, the likely conclusion is that 
the observed non-classical rotational inertia in $^4$He is not caused by supersolidity (for a review, see Ref.~\cite{boninsegni12}). In recent years, renewed interest in the subject has been triggered by a number of experiments with ultracold gases, in particular with Bose-Einstein
condensates in driven single or double cavities~\cite{baumann10,leonard17} or in the presence of spin-orbit coupling~\cite{li17}.  In both cases, the period of the 
density profile is set externally, either by the wave vector of the cavity photons or the momentum transfer associated with the Raman coupling 
between two internal states, which leads to a density modulation along a single direction. More recently, supersolid phases with an 
interaction-generated density modulation along the axial direction of a cigar-shaped trap have been realized with dipolar gases in a regime 
where the dipolar length $\ell_d$ is of the same order as the short-range scattering length $a_s$~\cite{boettcher19,tanzi19,chomaz19}. 
These supersolids are generically incommensurate, with many atoms per unit cell. In particular, there is typically a large superfluid fraction, 
which allows to observe signatures of supersolidity more easily.

In the present work, we analyze the spectrum of hydrodynamic and Goldstone modes for a general class of supersolids that 
exhibit a mass-density wave along a single direction. They may be thought of as a superfluid version of a classical smectic-A 
liquid crystal~\cite{chaikin95}. As emphasized by Martin {\it et al.}~\cite{martin72}, the structure of long-wavelength and low-energy excitations in any 
thermodynamic phase is completely described in terms of conserved variables and broken symmetries. 
The superfluid smectic, where two separate symmetries --- gauge invariance and translational invariance --- 
are spontaneously broken, is thus expected to exhibit an excitation spectrum with two separate Goldstone modes. 
This statement is not obvious, however, because Goldstone modes may be redundant.  As discussed by Watanabe and 
Murayama~\cite{watanabe13}, this happens, e.g., in the case of a vortex lattice, where only a single Goldstone mode is present
even though both gauge and translational invariance are broken. Physically, this arises from the fact that the deformation
field of the vortex lattice is not an independent variable but is rigidly coupled to the superfluid velocity. On a formal level, for a genuine supersolid, the Noether current densities for broken gauge and translation 
invariance must be independent. This requirement is automatically obeyed in supersolid phases with Galilean invariance~\cite{watanabe12}.
In addition, it also holds for the stripe phase of a spin-orbit coupled BEC as a consequence of the spin symmetry of this state,
despite the fact that Galilean invariance is broken\footnote{In the presence of spin-orbit coupling, the Galilean-invariant identity $T^{0i} = m j^i$ that links the generator of translations to the particle current receives additional corrections that involve the spin-projection $s_z$ of the two-component Bose gas~\cite{stringari17}. Since the smectic stripe phase is spin-balanced~\cite{li13}, these corrections do not enter.}.
 
For the superfluid smectic, broken translation invariance along a single direction implies that the superfluid mass 
density tensor is anisotropic and there is a finite normal fluid density for longitudinal motion even at zero temperature.
Besides the standard bulk sound mode, this gives rise to a separate second-sound type mode whose velocity is set by a 
combination of the layer compression modulus and the superfluid fraction. This is different from the individual cases of a 
smectic phase, where a secondary sound mode vanishes for propagation both along or perpendicular to the layer~\cite{chaikin95}, 
or a homogeneous superfluid, where second sound is an entropy wave that becomes ill-defined at low temperatures.

This paper is structured as follows: In Sec.~\ref{sec:2}, we determine the spectrum of longitudinal hydrodynamic 
and Goldstone modes of a superfluid smectic with both Galilean and time-reversal invariance.
The resulting first and second sound velocities turn out to depend only on three thermodynamic parameters, 
which are the bulk and layer compression modulus together with the superfluid fraction. In particular, it is shown 
that the second sound mode derives from a combination of two diffusive modes in the normal smectic, 
which are the heat diffusion mode and a characteristic permeation mode that describes defect diffusion. 
The physical nature of the second propagating mode is discussed in detail in Sec.~\ref{sec:3},
where the connection to the classic Andreev-Lifshitz picture of supersolids in terms of a propagating 
defect density mode is made. Section~\ref{sec:4} contains a discussion of our results and their relevance to recent measurements of the excitation spectrum of supersolid phases 
realized with dipolar gases. There are two appendices, one on the Leggett bound for the 
superfluid fraction in superfluids with an inhomogeneous density profile and a second one on the hydrodynamic modes 
transverse to the direction of spatial order. This allows to connect our results to earlier work 
by Radzihovsky and Vishwanath on superfluid liquid crystal 
states in imbalanced Fermi superfluids with either Larkin-Ovchinnikov or Fulde-Ferrell type order along a single direction~\cite{radzihovsky09,radzihovsky11}.  
 
\section{Hydrodynamic and Goldstone modes in a superfluid smectic}\label{sec:2}

\subsection{Hydrodynamics of a smectic-A liquid crystal}\label{sec:2a}

In order to elucidate the similarities and differences between standard liquid crystals and the superfluid version of the smectic phase considered in this paper, we start with the case where no superfluidity is present. To simplify the discussion while still keeping the essential physics of superfluid smectic phases, 
we consider a two-dimensional situation where the smectic order shows up as a weak periodic modulation 
\begin{equation}
n_{\rm eq}({\bf r})  =\bar{n}+\sum_{l=1}^{\infty} n_l \cos{(lq_0y)} \approx  \bar{n}+ n_1 \cos{(q_0y)}+\ldots 
\label{eq:smectic}
\end{equation}
of the density along the $y$-direction with a fundamental reciprocal lattice vector $q_0$.  
For a non-vanishing Fourier component $n_1\ne 0$ in Eq.~\eqref{eq:smectic}, translation invariance along $y$ is broken. The associated new hydrodynamic variable is a scalar field $u(x,y)$ that is called the layer phase~\cite{chaikin95}. It is defined by considering deviations from the equilibrium density~\eqref{eq:smectic} of the form
\begin{equation}
n(x,y) = \bar{n}+n_1 \cos{[q_0y-q_0u(x,y)]} \, .
\label{eq:layerphase}
\end{equation} 
At the level of a hydrodynamic description, there are four conserved quantities, which are particle number, the two-dimensional momentum 
as well as energy. Combined with the single symmetry-breaking variable $u$, there must be five hydrodynamic modes~\cite{martin72}. Only 
one of them is a Goldstone mode, which counts twice in a hydrodynamic count because it is necessarily a propagating mode.
As found by Martin {\it et al.}~\cite{martin72}, the Goldstone mode of a smectic-A liquid crystal 
is a transverse sound mode with a frequency $\omega_t(\mathbf{q})\simeq\sqrt{B/\rho q^2}\, q_xq_y\sim \sin{\psi}\cos{\psi}$ that
depends on the angle $\psi$ between the wave vector $\mathbf{q}$ and the direction of density order. Here, $\rho$ is the total equilibrium mass density and $B$ the layer compression modulus. It is defined by the elastic contribution $f_{\rm el}=B\, (u')^2/2+\ldots$ 
to the free energy density associated with small longitudinal distortions $u'=\partial_y u$ of the smectic order~\cite{chaikin95}. The second propagating mode is a bulk sound mode $\omega = \pm c_l q$ whose velocity $c_l(\psi)$ has only a weak dependence on the angle $\psi$~\cite{liao73}. 
In particular, for longitudinal propagation, its velocity 
\begin{equation}
c_l^2(\psi=0) = \frac{K+B}{\rho} 
\label{eq:velocity}
\end{equation} 
is determined by the sum of the (isentropic) bulk modulus $K = \rho~\partial p/\partial \rho\big|_{s,u'}$ 
and the layer compression modulus $B$~\cite{martin72}. For weak modulations of the density $n_1\ll \bar{n}$, 
the bulk modulus dominates and thus the sound velocity is essentially that of a fluid phase. The last remaining mode in 
addition to the Goldstone mode and the sound mode describes heat diffusion.

Consider now the special case in which the wave vector $\mathbf{q}$ is directed either along or perpendicular to the $y$-direction.
Here, due to the peculiar angular dependence $c_t(\psi)\sim  \sin{\psi}\cos{\psi}$ of the transverse sound velocity, the Goldstone 
mode is absent. By mode counting, there must then be three diffusive modes in addition to the propagating 
bulk sound mode. The first one is the heat diffusion mode that is present at arbitrary values of the angle $\psi$. 
The second one is a transverse momentum diffusion mode with frequency $\omega=-i\nu\,q^2$, where $\nu$
is a kinematic viscosity~\cite{chaikin95}. The third mode with frequency $\omega=-iD_p\,q^2$ is special to smectic-A liquid crystals and is called the permeation mode~\cite{chaikin95}. It describes a diffusive process in which particles are exchanged between adjacent layers without changing the average periodic structure. The associated diffusion constant $D_p=\zeta B$ is determined by  the layer compression modulus $B$  and a dissipative coefficient $\zeta$. 
The permeation mode may be viewed as an analog of vacancy diffusion, a process that gives rise to an independent hydrodynamic mode 
in any crystal~\cite{martin72}. As will be shown below, it is precisely the permeation mode in combination with the heat diffusion mode that 
turns into the Goldstone mode of the superfluid smectic phase, where exchange between the layers occurs in a reversible manner by 
non-dissipative, propagating mass currents.

\subsection{Longitudinal modes of a superfluid smectic-A phase}

For a description of the low-energy excitations of a superfluid smectic phase, the presence of superfluidity needs to be 
accounted for on a thermodynamic level by expressing the differential of the entropy density $s$ 
\begin{equation}
Tds=d\varepsilon - (\mu/m)d\rho -\mathbf{v}_n d\mathbf{g} - \mathbf{h}\, d(\nabla u) - \mathbf{j}_s\, d \mathbf{v}_s  
\label{eq:Gibbs}
\end{equation} 
as a function of the conserved variables energy density $\varepsilon$, mass density $\rho$, and momentum density 
$\mathbf{g}= {\underline{\rho_n}} \mathbf{v}_n + {\underline{\rho_s}} \mathbf{v}_s$ 
together with the gradient $\nabla u$ of the layer phase and the superfluid velocity $\mathbf{v}_s$, which characterize 
the two broken symmetries. 
Both are $U(1)$ symmetries 
and may therefore be derived from compact angular variables. In the superfluid, this is the standard phase $\theta_{\rm sf}\in (-\pi,\pi]$
whose gradient determines the superfluid velocity $\mathbf{v}_s=(\hbar/m) \nabla\theta_{\rm sf}$. Concerning the smectic order,
Eq.~\eqref{eq:layerphase} shows  that $q_0u(x,,y)$ and $q_0u(x,y)+2\pi$ give rise to identical distortions. 
Density fluctuations in the smectic are thus described by a different angle  
$\theta_{\rm sm}\in (-\pi,\pi]$ such that $\nabla u=(1/q_0)\nabla\theta_{\rm sm}$.
The formal equivalence of the order in smectic-A liquid crystals and in superfluids was in fact first realized by de~Gennes~\cite{degennes72}\footnote{Note that the associated angles $\theta_{\rm sm}$ and $\theta_{\rm sf}$ transform in an opposite manner
under time reversal: $\theta_{\rm sm}$ is a true scalar while $\theta_{\rm sf}$ is a pseudoscalar.}. 
The thermodynamic field 
\begin{equation}
\mathbf{h}= \frac{\partial f_{\rm el}}{\partial(\nabla u)}\bigg|_{T, A, N, v_n, v_s} = B u'\, \mathbf{e}_y - K_1\partial_x^3 u\, \mathbf{e}_x +\ldots
\label{eq:elastic}
\end{equation} 
conjugate to the gradient $\nabla u$ of the layer phase determines the elastic free energy of the smectic. Here and in the remainder 
of the paper, we shall use the notation $u' = \partial_y u$ for the derivative of the layer phase in the direction of the periodic modulation. 
Note that for longitudinal modes, only the layer compression modulus $B$ plays a role. A second Gaussian-curvature type elasticity 
appears in Eq.~\eqref{eq:elastic} for excitations with a finite component $q_x$ of the wave vector parallel to the layers. It involves
the splay elastic constant $K_1$~\cite{chaikin95}, which is relevant for the dispersion of Goldstone modes in
a Larkin-Ovchinnikov phase of imbalanced Fermi superfluids as discussed by Radzihovsky and 
Vishwanath~\cite{radzihovsky09,radzihovsky11}, see the discussion in~\ref{app:FFLO}.

The conjugate variable to the momentum density $\mathbf{g}$ is the normal velocity $\mathbf{v}_n$, which also appears 
in the superfluid mass current density $\mathbf{j}_s={\underline{\rho_s}} (\mathbf{v}_s-\mathbf{v}_n)$. The latter relation 
follows from the thermodynamic derivative ${\bf j}_s = \frac{\partial f}{\partial {\bf v}_s}\big|_{T, A, N, v_n, \nabla u}$, 
where the contribution to the free energy $f \sim - \frac{1}{2} \rho v_n^2 + \frac{1}{2} ({\bf v}_s - {\bf v}_n)^T {\underline{\rho_s}}  ({\bf v}_s - {\bf v}_n)$ 
is dictated by Galilean invariance~\cite{son05,yoo10}. Quite generally, for superfluids with an underlying periodic structure, the normal velocity $\mathbf{v}_n=\partial_t \mathbf{u}$ is 
determined by the time derivative of the displacement field  $\mathbf{u}$. This relation --- which is valid at the linearized level around 
equilibrium and is thus sufficient for the derivation of the hydrodynamic modes --- has been derived in general form by Son~\cite{son05}
as a consequence of Galilean invariance. In particular, for hydrodynamic modes with wave vector $\mathbf{q}$ along the $y$-direction, 
the associated normal velocity $v_{n,y}=\partial_t u$ is just the time derivative of the scalar layer-phase variable $u$. 
Similar to the elastic constants $B$ and $K_1$, the superfluid and normal mass density tensors ${\underline{\rho_s}}$ 
and ${\underline{\rho_n}}$, which are constrained by ${\underline{\rho_s}}+{\underline{\rho_n}}=\rho\,{\underline{1}}$, are
thermodynamic variables defined via $\mathbf{j}_s$ as the conjugate field to $\mathbf{v}_s$.  These relations are a straightforward 
tensor generali\-zation of those from standard two-fluid hydrodynamics. For translationally-invariant fluids,  
the normal fluid density ${\underline{\rho_n}} \sim T^{d+1}\,{\underline{1}}$ vanishes as the temperature approaches zero~\cite{fisher88}.
A different situation arises for a superfluid with smectic order, where the breaking of translation invariance 
along one of the directions gives rise to a finite value of the corresponding component $({\underline{\rho_n}})_{yy}\simeq \rho\cdot (n_1/\bar{n})^2$ 
of the normal fluid density even at zero temperature. 
A strict lower bound for $({\underline{\rho_n}})_{yy}$ follows from a variational argument due to Leggett~\cite{leggett70}, 
which is discussed in~\ref{app:leggett}. 

The complete set of hydrodynamic modes in a superfluid smectic phase follows from the equations of motion for the 
conserved densities together with the two variables that describe the underlying broken symmetries.
The latter are the superfluid velocity ${\bf v}_s$ and the gradient of the layer phase $\nabla u$, which are both longitudinal 
vectors and thus effectively scalar quantities.  Together with the dynamic equations for the particle density~$\rho$, the momentum 
density ${\bf g}$ and energy density $\varepsilon$, the six resulting equations of motion are given by
\begin{align}
&\partial_t \rho + \nabla \cdot {\bf g} = 0 \label{eq:C1} \\
&\partial_t g_i + \partial_j \pi_{ij} = 0 \label{eq:C2} \\
&\partial_t \varepsilon + \nabla \cdot {\bf j}^\varepsilon = 0 \label{eq:C3} \\
&\partial_t (\nabla u) - \nabla v_{n,y} = 0 \label{eq:C4} \\
&\partial_t {\bf v}_s + \nabla \mu/ m= 0 \label{eq:C5} .
\end{align}
The first three equations~\eqref{eq:C1}-\eqref{eq:C3} are continuity equations that link the time derivatives of the densities to the 
divergences of the momentum density ${\bf g}$, the stress tensor $\pi_{ij}$, and the energy current~${\bf j}^\varepsilon$, respectively. 
As already discussed above, Eq.~\eqref{eq:C4} expresses that a constant shift along the direction of 
smectic order changes the layer phase by a constant\footnote{If irreversible effects are included, the right-hand side in this equation no longer 
vanishes and contains a contribution $\zeta\nabla^2 h$, where $\zeta$ is the dissipative coefficient that enters the diffusion constant $D_p=\zeta B$ 
of the permeation mode~\cite{chaikin95}.}. 
Finally, Eq.~\eqref{eq:C5} is the Josephson equation (neglecting a quadratic term in the velocities) that describes the dynamics of the superfluid phase.

From the differential of the entropy~\eqref{eq:Gibbs} and the dynamic equations~\eqref{eq:C1}-\eqref{eq:C5}, we obtain 
an expression for the material derivative $T (\partial_t s\!+\!{\bf v}_n\!\cdot\!\nabla\!s)$ of the entropy density that depends on
spatial gradients $\nabla T$, $\nabla \mu$, $\partial_i v_{n,j}$ and $\nabla \cdot {\bf j}_s$ of the thermodynamic forces. For the 
inviscid fluid considered here there is no entropy production, which implies a particular series of constitutive relations that link
the currents and the thermodynamic forces. To leading order in the velocities, these constitutive relations read:
\begin{align}
{\bf g} &= \rho {\bf v}_n + {\bf j}_s \label{eq:const1} \\
\pi_{ji} &= p \delta_{ij} - (h_{i} \delta_{j,y}) \\
j_i^\varepsilon &= (\varepsilon + p) v_{n,i} + \mu j_{s,i}/m . \label{eq:jE}
\end{align}
Compared to a simple fluid, at this level the superfluid order modifies the particle and energy current, while the smectic order 
modifies the stress tensor. In addition, as stated above, the thermodynamic forces ${\bf j}_s$ and~${\bf h}$ are linked to the velocities by
$\mathbf{j}_s = {\underline{\rho_s}} (\mathbf{v}_s-\mathbf{v}_n)$ and $\mathbf{h} = B\partial_y u\, \mathbf{e}_y$. 

The linearized hydrodynamic equations of motion are obtained by substituting the constitutive relations in the dynamic equations 
and expanding the thermodynamic forces to leading order in the hydrodynamic variables around equilibrium. 
For motion along the direction of the smectic order (here, the $y$-direction), the resulting equations 
only involve the $yy$ component of the superfluid mass density tensor, which we denote by $\rho_s=\rho -\rho_n$ in the following. 
In this configuration, the transverse momentum degree of freedom decouples and gives rise to a diffusion mode. 
For the remaining five degrees of freedom, we obtain the characteristic equation
\begin{align}
\begin{pmatrix}
- \omega/q & 1 & 0 & 0 & 0 \\
K/\rho & - \omega/q  & 0 & 0 & - B \\
0 &
\tilde{s} T \frac{\rho_s}{\rho_n}
& - \omega/q  & 
- \rho \tilde{s} T \frac{\rho_s}{\rho_n}  & 0 \\
K/\rho^2 & 0 & - \tilde{s}/\rho c_V & - \omega/q  & 0 \\
0 &
1/\rho_n & 
0 &
- \rho_s/\rho_n & - \omega/q 
\end{pmatrix}
\begin{pmatrix}
\delta \rho \\
g_L \\
\delta q \\
v_s \\
u'
\end{pmatrix}
&= 0 . \label{eq:matrix}
\end{align}
Here, $\omega$ is the frequency of the mode and $q$ the associated longitudinal momentum. Moreover, we introduce a heat current density 
variable $\delta q = \delta \varepsilon + \frac{\varepsilon+p}{\rho} \delta \rho$ with $\tilde{s} = s/\rho$ the entropy per particle and mass, while
$c_V = T \frac{\partial\tilde{s}}{\partial T}\bigr|_{\rho}$ is the associated specific heat. Apart from the transverse momentum diffusion 
mode mentioned above, Eq.~\eqref{eq:matrix} contains another diffusive zero mode with eigenvector 
$(\delta \rho, g_L, \delta q, v_s, u') = (-B \rho/K, 0, - B c_V/\tilde{s}, 0, 1)$. In the absence of superfluidity, this mode splits into two separate 
diffusion modes, one that describes heat diffusion, and one permeation mode that involves an interchange between the layer phase and the 
particle density. In the superfluid smectic phase, only the above combination remains, 
while an ortho\-gonal complement will couple to the superfluid velocity and give rise to a propagating sound mode.

The determinant of Eq.~\eqref{eq:matrix} that determines the propagating longitudinal hydrodynamic modes reads
\begin{align}
&\omega^4 + \omega^2 q^2 \biggl[- \frac{K}{\rho} - \frac{B}{\rho_n} - \frac{\tilde{s}^2 T}{c_V} \frac{\rho_s}{\rho_n}\biggr] + 
q^4 \biggl[\frac{B K}{\rho^2} \frac{\rho_s}{\rho_n}+ \frac{\tilde{s}^2 T}{c_V} \frac{\rho_s}{\rho_n} \frac{K-B}{\rho} \biggr] = 0 . \label{eq:characteristic}
\end{align}
Neglecting the terms of order $\tilde{s}^2 T/c_V$ at low temperatures,  we obtain two undamped propagating
modes $\omega=\pm c_{1,2}q$ with velocities 
 \begin{equation}
c_{1,2}^2=\frac{K}{2\rho} +\frac{B}{2\rho_n} \pm \frac{1}{2}\biggl[ \left(\frac{K}{\rho} +\frac{B}{\rho_n}\right)^2 - 4f_s\frac{KB}{\rho\rho_n}\biggr]^{1/2}\, .
\label{eq:velocity1}
\end{equation} 
The physics behind these two longitudinal modes and in particular also the associated eigenvectors will be discussed in 
detail in the following section. We note that the result~\eqref{eq:velocity1} turns out to be equivalent to a result 
obtained by Yoo and Dorsey~\cite{yoo10} for the longitudinal hydrodynamic modes of a crystalline supersolid. 
Indeed, Eq.~\eqref{eq:velocity1} agrees with their Eqs.~(48) and~(49) if we identify the parameters $1/\chi \to K$ with the bulk compression modulus 
and $\lambda \to B$ with the layer compression modulus. Moreover, we do not include a strain-density coupling and thus $\gamma = 0$ in the 
notation of Ref.~\cite{yoo10}. However, heat currents are neglected in Ref.~\cite{yoo10} and therefore the contribution ${\cal O}(\tilde{s}^2 T/c_V)$ in 
Eq.~\eqref{eq:characteristic} is absent. As discussed in~\ref{app:FFLO}, this is of relevance if one considers the hydrodynamic modes
parallel to the layers, which involve a conventional entropic second sound mode. In general, therefore, the hydrodynamic modes of a superfluid 
smectic phase differ from those of supersolids in which translation invariance is broken in all directions --- it is only in the specific case of 
purely longitudinal propagation that both systems behave in a similar manner. An example for this is provided by the incommensurate supersolid 
phase of a two-dimensional Bose gas with interactions described by a soft disc potential, whose longitudinal modes have been determined numerically~\cite{saccani12,macri13}. We emphasize that Eq.~\eqref{eq:velocity1} for the longitudinal sound velocities is not restricted to a smectic-A phase but also applies to the longitudinal modes in a two- or three-dimensional crystal, as discussed, e.g., by Yoo and Dorsey~\cite{yoo10}.

\section{Defect density propagation and the limit of fourth sound}\label{sec:3}

For a better understanding of the physics underlying the two propagating modes found in Eq.~\eqref{eq:velocity1} and in particular the 
connection to the classic picture of supersolids in terms of wave-like propagation of defects proposed by Andreev and Lifshitz~\cite{andreev69}, 
it is instructive to rederive the longitudinal modes~\eqref{eq:velocity1} and the associated eigenvectors with a slightly different set of variables 
introduced by Yoo and Dorsey~\cite{yoo10}. They decompose small fluctuations of the mass density
\begin{equation}
\delta\rho=-\rho\, u' +\delta\rho_{\triangle}
\label{eq:defect}
\end{equation} 
into a contribution $-\rho\, u'$ associated with deformations of the periodic structure and an additional defect density $\rho_{\triangle}$. 
This separates the density variation of a defect free crystal, for which a change in density is tied to the divergence of the deformation 
field, from the additional density change associated with the motion of defects or interstitials. The defect density obeys a continuity equation 
\begin{align}
 \partial_t \delta \rho_{\triangle}=- \partial_y\, \rho_s (v_s\!-\!v_n) 
 \end{align}
whose conserved current $g_\Delta = \rho_s (v_s\!-\!v_n)$ is just the Galilean-invariant superfluid mass-current density, 
determined by the counterflow between the superfluid and the normal velocity $v_n=\partial_t u$. The second time derivative 
of the defect density is coupled to the strain field variable $u'$ according to  
\begin{align}
\partial_t^2 \delta \rho_\triangle &= \rho_s \partial_y^2 (\mu/m) + \rho_s \partial_t^2 u'  \, . \label{eq:defectdensity}
\end{align}
In a situation where the lattice is almost rigid, the contribution that involves the layer phase variable $u'$ may be neglected. 
As a result, the defect density exhibits wave-like propagation with a velocity given by $c_4^2 = f_s \, (K/\rho)$. This is analogous 
to fourth sound of superfluid $^4$He in narrow capillaries, where the normal fluid component is pinned by the walls.
It describes the oscillation of the superfluid with no motion of the lattice, a limit which is perfectly realized in the 
superfluid phase of bosons in an optical lattice~\cite{greiner02}, where the periodic modulation of the density is 
externally imposed and not caused by interactions.    
For the superfluid smectic, this limit is reached when the layer compression modulus contribution $B/\rho_n\gg K/\rho$ in Eq.~\eqref{eq:velocity1}
dominates that of the bulk. In general however, as Eq.~\eqref{eq:defectdensity} shows, the defect density and the longitudinal strain $u'$ are coupled. 
An explicit result for the eigenmodes of the superfluid smectic phase thus requires to simultaneously 
solve the equation for $\delta \rho_\triangle$ and for $u'$, which reads
\begin{align}
\rho_n \partial_t^2 u' &= \partial_y^2 [- p + B u' + \rho_s (\mu/m)] . \label{eq:eom2}
\end{align}
The solution of the coupled equations~\eqref{eq:defectdensity} and~\eqref{eq:eom2} 
does of course reproduce the result~\eqref{eq:velocity1} above. The associated 
dimensionless eigenvectors are 
\begin{align}
\begin{pmatrix}\delta \rho_\triangle/\rho \\[1ex] u'\end{pmatrix}_1 = 
\begin{pmatrix} c_2^2/(K/\rho) \\[1ex] 1\end{pmatrix}
\end{align}
for the first sound mode with speed $c_1$ and 
\begin{align}
\begin{pmatrix}\delta \rho_\triangle/\rho \\[1ex] u'\end{pmatrix}_2 = 
\begin{pmatrix} c_1^2/(K/\rho) \\[1ex] 1\end{pmatrix}
\end{align}
for the second sound mode with speed $c_2$. Specifically, for an almost rigid lattice with $B/\rho_n\gg K/\rho$,  
the velocities reduce to $c_1^2 = B/\rho_n + f_n K/\rho$ and $c_2^2 = f_s K/\rho$ with $c_1\gg c_2$. 
In this limit, therefore, second sound is essentially a defect density mode with no involvement of the lattice. 
Recall that this mode derives from the diffusive permeation mode of a smectic, which describes particle diffusion
without a change in the periodic structure. By contrast, the eigenvector $(\delta \rho_\triangle/\rho, u')_1 = (f_s,1)$ 
for first sound  in this limit involves the defect density with weight $f_s$. In standard supersolids, where $f_s$ is expected 
to be small compared to one, this mode predominantly involves  the strain field, i.e., it describes the motion of the lattice.

A rather different situation arises in the opposite limit of a small normal fraction $f_n\ll 1$ on top of a dominant homogeneous superfluid.  
Formally, in the limit $f_n\simeq (n_1/\bar{n})^2 \to 0$ of a vanishing density modulation,
the contribution $B/\rho_n$ in Eq.~\eqref{eq:velocity1} appears to diverge. This is not the case, however, since  the elastic constant 
$B$ approaches zero as well. The way it does has been discussed in the context of the nematic-to-{smectic\nobreakdash-A} transition of 
normal liquid crystals~\cite{chaikin95}: Within a mean-field approximation, the layer compression modulus $B\sim |n_1|^2$ vanishes like 
the square of the order parameter $n_1$. As a result, the ratio $B/\rho_n$ turns out to be finite in the limit $n_1\to 0$ where the smectic 
order disappears. Obviously, this result cannot remain valid beyond mean-field theory since it would predict a finite second sound velocity $c_2$ 
right at the transition to the fluid phase. The layer compression modulus $B\sim |n_1|^x$ must therefore vanish with an exponent $x>2$. 
In fact, as was shown by Grinstein and Pelcovits~\cite{grinstein81}, the renormalized value $B_{\rm ren}$ vanishes at long distances even in the smectic 
phase with $n_1 \ne 0$ due to anharmonic corrections to  the linear elastic continuum model used above. In the following, this complication 
will be ignored. The ratio $B/\rho_n$ is therefore a thermodynamic parameter which is finite in the superfluid smectic, but vanishes in the 
homogeneous superfluid where translation invariance is not broken.  In particular, in the limit $K/\rho\gg B/\rho_n$ of a weak density 
modulation, the velocities~\eqref{eq:velocity1} approach $c_1^2=(K+B)/\rho$ and $c_2^2=B/\rho_n$. 
The velocity of the compression mode is thus unchanged compared to that in the normal phase. 
In terms of the variables $\delta \rho_\triangle/\rho$ and $u'$, the eigenvector associated with first sound is 
dominated by the layer phase variable with a negligible contribution from the defect density. Due to $u'\simeq -\delta\rho/\rho$, 
the periodic structure of the smectic therefore adiabatically follows the density fluctuations $\delta\rho$ in this mode, 
which describes oscillations of the lattice. The second sound mode, by contrast, 
whose velocity $B/\rho_n$ is determined by the ratio of the layer compression modulus $B$ and the normal fluid density,
involves both an oscillation in the longitudinal strain field as well as the defect density with essentially equal magnitude. In physical terms, it describes a wave-like propagation of particles in addition to that associated with variations in the smectic 
lattice structure, replacing the diffusive permeation mode of a normal smectic phase.

\section{Summary and experimental implications of the results}\label{sec:4}

We have derived the spectrum of hydrodynamic modes in supersolids that exhibit a density modulation 
along a single direction only, with an arbitrary number of particles in a unit cell.  This state may be viewed as a superfluid version 
of a  classical smectic-A liquid crystal. It has a highly anisotropic spectrum of modes that is entirely determined by the 
number of broken symmetries and a few thermodynamic parameters.  In particular, the longitudinal excitations exhibit 
a second sound like mode that remains well-defined even at zero temperature. An analytical result [Eq.~\eqref{eq:velocity1}] 
has been derived for the sound velocities, which only contains the effective layer compression modulus and the superfluid fraction as the two low-energy parameters associated with 
broken translation and gauge invariance.
  
As mentioned in the introduction, a variety of phases where superfluidity coexists with periodic spatial order
have been observed in the context of ultracold gases in recent years. Among those, supersolids realized in driven cavities~\cite{baumann10,leonard17} are special since the cavity field gives rise to an infinite-range effective interaction between the atoms which leads to an incompressible system~\cite{piazza13}. The resulting Goldstone modes for the particular case of a double cavity, where a continuous form of translation symmetry breaking appears, have been 
discussed in Ref.~\cite{lang17}. Our present results, in turn, apply to compressible supersolid phases 
with short-range interactions.  In addition, translation symmetry is  assumed to be broken only along a single direction. The two examples where this type of symmetry breaking appears in experiments are the stripe phase of BECs in the presence of spin-orbit coupling and dipolar gases in a cigar-shaped trap.  

In the stripe phase of BECs with spin-orbit coupling, the superfluid exhibits a periodic density modulation along a 
single direction of the form assumed in Eq.~\eqref{eq:smectic}. 
Physically,  it arises from the momentum transfer associated with the Raman coupling between two internal states. In particular, the amplitude $n_1$ may be tuned by the strength $\Omega$ of the Raman coupling~\cite{li12}. 
The role of the layer phase variable $u$ is played by the relative phase $\phi$ between the complex coefficients
$C_1$ and $C_2$ in the Gross-Pitaevskii ansatz~\cite{li12}
\begin{equation}
\biggl(\begin{matrix} \psi_a\\ \psi_b \end{matrix}\biggr) =\sqrt{\frac{N}{V}}\biggr[ C_1 \biggl(\begin{matrix} \cos{\theta}\\ -\sin{\theta} \end{matrix}\biggr) e^{ik_1x} 
+ C_2 \biggl(\begin{matrix} \sin{\theta}\\ -\cos{\theta} \end{matrix}\biggr) e^{-ik_1x}\biggr] 
\label{eq:spinor}
\end{equation} 
for the spinor wave function, where $\theta$ is a variational parameter and $k_1$ sets the density modulation with wave vector $q_0=2k_1$. The spectrum of elementary excitations of the stripe phase has been determined by Li {\it et al.}~\cite{li13} within a Bogoliubov approach. 
For wave vectors along the direction of ordering, there are two gapless modes whose dispersion $\omega(q+q_0)=\omega(q)$ is
periodic in the associated Brillouin zone.  The upper one is a density mode with a velocity $c_1$ while the lower 
one is a spin excitation, again with a linear spectrum $\omega_2=c_2q$ at small values of the longitudinal wave vector $q$. Moreover, two separate sound modes exists also for a propagation parallel to the stripes~\cite{li13}. 
This differs from the superfluid smectic discussed in the present 
paper where for motion parallel to the stripes only the compression
mode with velocity $c_1$ survives at zero temperature, while a
second sound mode as an entropy wave at constant pressure is ill-defined
(see~\ref{app:FFLO}). It is an open problem to extend our analytical result~\eqref{eq:velocity1} for the two velocities in the supersolid phase of a single-component BEC to this two-component system and also to determine the effective layer compression modulus $B$ and the associated finite normal fluid density $\rho_n$ in terms of the microscopic parameters of spin-orbit coupled BECs~\cite{jian11,sanchezbaena20}.

An example where our results are of direct experimental relevance are dipolar gases in cigar-shaped traps. If the dipolar length $\ell_d$ is larger than the tunable scattering length $a_s$ associated 
with short-range interactions, they   
exhibit a supersolid phase with an interaction-driven density modulation along the weakly-confined direction~\cite{boettcher19,tanzi19,chomaz19}. In practice, these are highly inhomogeneous systems with only a few times $10^4$ atoms. The modulation of the density in the supersolid phase, which has a typical length of the unit cell of around $0.3\, \mu$m, splits the BEC into a small number of coherently coupled droplets.  Experimentally, a characteristic signature of the supersolid state in a trap compared to the standard BEC is the emergence of an additional collective mode at low energies~\cite{guo19,tanzi19b,natale19}. Specifically, as observed by Tanzi {\it et al.}~\cite{tanzi19b}, the axial breathing mode of a trapped BEC with frequency $\omega_B=\sqrt{5/2}\omega_y$ shifts towards higher frequencies beyond the transition to a state with finite density modulation. In addition, a new mode appears whose frequency goes down as the density contrast increases. This observation can be understood on a qualitative level within our hydrodynamic approach for a homogeneous system
by noting that the lowest value $q_{\rm min}\simeq 1/l_y$ of the longitudinal wave vector in the trap is set by the inverse of the axial confinement length $l_y$. As discussed in Sec.~\ref{sec:3}, the supersolid phase is characterized by a compression mode tied to an oscillation in the lattice structure and a lower-energy Goldstone 
mode that describes dissipationless transport of defects, independent of a change in the lattice constant. In the presence of a trap,
the splitting of the Bogoliubov-Anderson mode of a homogeneous BEC into two independent propagating modes in a supersolid phase
shows up as a bifurcation into a compressional mode at $\omega_1\simeq c_1/l_y$, the frequency of which is shifted upwards in
the supersolid phase because of the corresponding shift in the sound velocity through the additional contribution of the layer compression modulus $B$. 
In addition, a second mode appears at lower frequencies $\omega_2\simeq c_2/l_y$. It becomes increasingly soft upon 
entering more deeply into the supersolid phase since the velocity $c_2$ decreases with the superfluid fraction $f_s$. In particular, in the limit where the periodic structure is essentially rigid, it approaches $c_2=\sqrt{f_sK/\rho}$, which vanishes as the square root of the superfluid fraction $f_s\to 0$ near the superfluid-to-Mott-insulator like transition of the supersolid to a crystal of droplets that have no long-range phase coherence. A quantitative comparison between our continuum results and the experimental data is unfortunately not straightforward. It would require to properly describe the crossover from genuine excitations in the trap such as the breathing mode at $\omega_B=\sqrt{5/2}\omega_y$, which are completely independent of interactions, to the 
collective excitations studied in our present work. A trap analog of the true Goldstone mode, which is properly defined only in a homogeneous system, has also been seen in experiments by Guo {\it et al.}~\cite{guo19} 
in a small array of three droplets, where an in-phase dipole mode associated with the axial motion of the whole cloud coexists  with an out-of-phase mode at frequencies much smaller than that of the trap. In the latter, atoms are moving between the droplets at a fixed center-of-mass position, corresponding to a counterflow between the periodic lattice and a superfluid of defects on top, analogous to the mode at $\omega_2\simeq c_2/l_y$ discussed above.

A possible way to measure the sound velocities predicted in Eq.~\eqref{eq:velocity1} --- and thus to extract quantitative values for the layer compression modulus $B$ and the superfluid fraction $f_s$ 
even with dipolar gases in cigar-shaped traps --- is suggested by the experiments of 
Petter {\it et al.}~\cite{petter19}. By an extrapolation to small wavevectors, they allow to infer the presence 
of a sound-like excitation in the regime of a homogeneous superfluid with a linear dispersion $\omega=c_{1}\,q$, which in fact persists for 
wave vectors up to~$q = l_z^{-1} \sim 3.6 l_y^{-1}$~\cite{petter19}. Provided that the extrapolation into the linear regime is 
possible in the supersolid phase, our analytical results for the mode velocities in 
Eq.~\eqref{eq:velocity1} will allow to determine the three parameters involved in the thermodynamic description of the superfluid smectic\footnote{The bifurcation of the standard Bogoliubov mode $\omega_q=c_{1}\,q$ into two independent ones in the supersolid phase 
has been seen in numerical simulations based on a modified Gross-Pitaevskii equation, see Ref.~\cite{natale19}.}. 
Specifically, the bulk compression modulus $K$ is fixed by the velocity $c_1^2=K/\rho$ of first sound in the homogeneous
superfluid before the density wave instability. A measurement of the two velocities in the superfluid smectic will then 
uniquely determine the two remaining parameters $B/\rho_n$ and $f_s$, using for example the relations 
$c_1^2+c_2^2=K/\rho + B/\rho_n$ and $c_1^2c_2^2=f_s\cdot (K/\rho)\, (B/\rho_n)$. 
Beyond a direct measurement of the layer compression modulus $B$, this might allow to check the values of the superfluid fraction $f_s$ 
extracted from the contrast $C=(n_{\rm max}-n_{\rm min})/(n_{\rm max}+n_{\rm min})$ of the density profiles in Ref.~\cite{petter20} via the Leggett bound. 
Moreover, the recent observation of a normal phase with a
finite density modulation along the axial direction by Sohmen {\it et al.}~\cite{sohmen21}
in principle allows to determine the two parameters $B$ and $f_s$ in Eq. (16)   
from independent measurements in the normal and the superfluid smectic phase.
An open question in this context is whether the superfluid-to-supersolid transition is continuous or first order with a corresponding jump in the contrast, as suggested by the discontinuity in the phonon velocities of the Goldstone modes obtained within 
a Bogoliubov description~\cite{macri13} and a recent generalization of the classical 
Hansen-Verlet criterion for freezing to quantum fluids~\cite{hofmann20}.

In a more general context, an interesting challenge for future studies is to investigate the hydrodynamics of supersolid phases 
with a frozen, inhomogeneous density that is not periodic. Such a superglass phase has been found in numerical simulations of 
Bose fluids with strong short-range repulsion like $^4$He that are rapidly quenched to low temperatures
~\cite{boninsegni06}. They violate 
Galilean invariance and thus might realize a superthermal phase conjectured some time ago by Liu~\cite{liu78},
in which a finite temperature gradient can be sustained in a static, non-dissipative situation.  

\ack

It is a pleasure to acknowledge a number of helpful comments by A. T. Dorsey, A. J. Leggett, S. Moroz and L. Radzihovsky. This work is supported by Vetenskapsr\aa det (grant number 2020-04239) (JH).

\appendix

\section{Leggett bound on the superfluid fraction}\label{app:leggett}

Leggett derived an upper bound for the superfluid fraction $f_s$ in a ground state with broken gauge and translation
invariance, which only involves the microscopic density profile~\cite{leggett70}. 
In the special case of a purely one-dimensional configuration, Leggett's result states that the superfluid fraction
 \begin{equation}
f_s= \frac{\rho_s}{\rho}\leq \frac{b}{\bar{n}\int_0^b dy/n(y)} 
  \label{eq:Leggett}  
 \end{equation}
is bounded from above by an integral over a unit cell of the lattice (taken to be along
the $y$-direction as in Eq.~(\ref{eq:smectic})) with lattice constant $b=2\pi/q_0$ and average density $\bar{n}$. 
It is important to note that the bound does not rely on any 
commensurability condition: it applies both to a commensurate situation, where the product  $\bar{n}\,b\!=\!k$ 
of the average density and the lattice constant $b$ is an integer $k=1,2,\ldots$, or the generically incommensurate 
case associated with a weak mass-density wave, which is of relevance for the superfluid smectic. 
The bound becomes increasingly tight for densities that are strongly suppressed at intermediate points 
within a unit cell, as expected in a real crystal. In turn, 
superfluidity is favored if the density exhibits only small fluctuations around its average $\bar{n}$.  Of course, 
the bound~(\ref{eq:Leggett}) does not provide a sufficient criterion for superfluidity in a state with broken translation invariance: 
a finite value of the bound is still compatible with no superfluidity at all. What it shows, however, is that a ground state of 
bosons with non-uniform density necessarily has a finite normal fluid fraction.

Within a Gross-Pitaevskii description of the superfluid smectic phase (see, for example, Ref.~\cite{tanzi19b}),
it is assumed that the one-particle density operator 
\mbox{$\hat{\rho}_1=\sum_{\alpha} \lambda_{\alpha}^{(1)} \vert \psi_{\alpha}\rangle \langle \psi_{\alpha}\vert$} 
is dominated by a single macroscopic eigenvalue $\lambda_0^{(1)}\simeq N$. In the regime, where the 
ground state exhibits a weak density wave, the associated eigenfunction 
$\langle\mathbf{x}  \vert \psi_{0}\rangle\sim 1 + \delta \cos{(q_0y)} +\ldots$
involves a small admixture of order $|\delta|\ll 1$, which breaks translation symmetry along the $y$-direction 
(in Ref.~\cite{petter20}, this is called the sine ansatz). The resulting equilibrium density 
\begin{equation}
n_{\rm eq}(\mathbf{x})  = \langle\mathbf{x}\vert\hat{\rho}_1\vert\mathbf{x}\rangle\,\to\,\frac{\bar{n}}{(1+\delta^2/2)}\left[ 1+ \delta \cos{(q_0y)}\right]^2 
\label{eq:GP-density}
\end{equation}
is then of the form assumed in Eq.~(\ref{eq:smectic}) with $n_1/\bar{n}\simeq 2\delta$ to linear order in $\delta$. 
For $|\delta|\geq 1$, the density~(\ref{eq:GP-density}) vanishes quadratically at either one (for $|\delta|=1$) or two different points 
within the unit cell, which leads to a divergent integral in the denominator of Eq.~(\ref{eq:Leggett}).
This is revealed by the special form 
 \begin{equation}
f_s\leq \frac{(1-\delta^2)^{3/2}}{1+\delta^2/2}\to 1-2\delta^2 \; {\rm for} \; \delta\to 0
  \label{eq:bound}  
 \end{equation}
of the Leggett bound for the Gross-Pitaevskii state, which becomes increasingly tight for a large density modulation and is ill-defined for \nobreak $|\delta|> 1$. 
As pointed out in the main text, the Leggett bound implies that the normal fluid fraction $f_n\geq 2\delta^2 +\ldots$ in a state of the form~\eqref{eq:GP-density}
is bounded from below by a finite value even at zero temperature. Unfortunately, the precise numerical
factor connecting $f_n$ with the square of the density modulation is not determined by this variational argument.

Finally, we emphasize that the use of the Leggett bound to extract a finite normal fluid density from a Gross-Pitaevskii ansatz  for the supersolid ground state requires to take into account excitations beyond the Gross-Pitaevskii description. 
Indeed, the bound~\eqref{eq:Leggett} relies on applying a finite total twist $\Theta=\int_y\partial_y\theta_{\rm sf}(y)$ of the 
local phase $\theta_{\rm sf}(y)$ of the eigenfunction $\langle\mathbf{x}  \vert \psi_{0}\rangle$ and allowing the system 
to adjust the functional form of $\theta_{\rm sf}(y)$ to minimize the energy. The minimum is achieved by putting in 
the imposed phase change $\Theta$ in regions of small density, which leads to the bound~(\ref{eq:Leggett}). 
In the context of a Gross-Pitaevskii description of supersolid phases, the problem of properly defining a normal fluid fraction 
has been addressed by Josserand {\it et al.}~\cite{josserand07a, josserand07b}. They suggest to decompose the amplitude and 
phase of the wave function into slowly- and rapidly-varying components and eliminate the latter. In practice, the elimination of the fast degrees of freedom, varying on the scale of the lattice constant, cannot be performed 
in explicit form. It is thus not clear whether this procedure is able to describe supersolids with finite values of $f_n$ and eventually recover a normal solid phase with $f_n=1$.

\section{Transverse modes of a superfluid smectic and the relation to superfluids with unidirectional Fulde-Ferrell or Larkin-Ovchinnikov order}\label{app:FFLO}

In this appendix, we provide results for the hydrodynamic modes of the smectic~A superfluid beyond the purely longitudinal situation discussed in Sec.~\ref{sec:2}. We begin by collecting the dynamical equations for the hydrodynamic variables $(\delta \rho, g_L, g_T, \delta \varepsilon, v_s, \nabla u)$, Eqs.~\eqref{eq:C1}-\eqref{eq:C5}, written in Fourier space for arbitrary angles
 $\psi$ between the direction of propagation and the direction of smectic order. For this, it is helpful to decompose the current into 
 a longitudinal and a transverse part defined by
\begin{align}
g_L &= \frac{q_x}{q} g_x + \frac{q_y}{q} g_y \\
g_T &= \frac{q_x}{q} g_y - \frac{q_y}{q} g_x .
\end{align}
In terms of the longitudinal current, the continuity equation~\eqref{eq:C1} reads:
\begin{align}
- \frac{\omega}{q} \delta \rho + g_L &= 0 . \label{eq:density1}
\end{align}
To rewrite expression~\eqref{eq:C2} for the longitudinal current, we introduce the longitudinal and 
transverse part of the conjugate field ${\bf h}$, which read with our choice~\eqref{eq:elastic} of the elastic free energy
\begin{align}
h_{u,L} &= u' B \cos^2 \psi \\
h_{u,T} &= u' B \cos \psi \sin \psi .
\end{align}
Moreover, in expanding the pressure in terms of the hydrodynamic variables, we neglect its dependence on entropy, superfluid velocity, 
and layer phase gradient\footnote{
The fact that to leading order in the velocities, the pressure will not depend on the superfluid velocity
or the layer phase gradient follows from our definition of ${\bf h}$ and ${\bf j}_s$ in combination with Maxwell relations that link ${\bf h}$ 
and the pressure, which in turn can be derived from the thermodynamic differential 
$d\tilde{\varepsilon} = T d\tilde{s} - p d(1/\rho) + {\bf v}_n d({\bf j}/\rho) +({\bf j}_s/\rho) d{\bf v}_s + ({\bf h}_u/\rho) d(\nabla u)$ for the energy 
per particle $\tilde{\varepsilon} = \varepsilon/\rho$.}. 
We obtain
\begin{align}
&- \frac{\omega}{q} g_L + p - h_{u,L} \cos \psi \nonumber \\
&\qquad = - \frac{\omega}{q} g_L + \bigl[K \delta \rho - B \cos^3 \psi \, u'\bigr] = 0 . \label{eq:appCjL}
\end{align}
Likewise, Eq.~\eqref{eq:C2} for the transverse current becomes
\begin{align}
&- \frac{\omega}{q} g_T - h_{u,L} \sin \psi \nonumber \\
&\qquad = - \frac{\omega}{q} g_T - B \cos^2 \psi \sin \psi \, u'= 0 . \label{eq:appCjT}
\end{align}
In order to derive an expression for the energy density~\eqref{eq:C3}, we use the continuity equation~\eqref{eq:density1}, 
the constitutive relations for the current~\eqref{eq:const1}, as well as the expression for the pressure,
\begin{align}
- p &= \varepsilon - s T - \mu \rho - {\bf j} \cdot {\bf v}_n , \label{eq:defp}
\end{align}
which has the thermodynamic differential
\begin{align}
dp &= s dT + \rho d\mu + {\bf j} \cdot d{\bf v}_n - {\bf j}_s \cdot d{\bf v}_s - {\bf h} \cdot d(\nabla u) . \label{eq:defdp}
\end{align}
In addition, substitute the decomposition of the superfluid current
\begin{align}
j_{s,L} &= - g_L \biggl[\frac{\rho^y_s}{\rho^y_n} \cos^2 \psi + \frac{\rho^x_s}{\rho^x_n} \sin^2 \psi\biggr] \nonumber \\
&\qquad - g_T \cos \psi \sin \psi \biggl[\frac{\rho^y_s}{\rho^y_n} - \frac{\rho^x_s}{\rho^x_n}\biggr]
+ \rho v_s \biggl[\frac{\rho^y_s}{\rho^y_n} \cos^2 \psi + \frac{\rho^x_s}{\rho^x_n} \sin^2 \psi\biggr] 
\\
j_{s,T} &= g_L \cos \psi \sin \psi \biggl[\frac{\rho^x_s}{\rho^x_n} - \frac{\rho^y_s}{\rho^y_n}\biggr] \nonumber \\
&\qquad- g_T \biggl[\frac{\rho^y_s}{\rho^y_n} \sin^2 \psi + \frac{\rho^x_s}{\rho^x_n} \cos^2 \psi\biggr]
+ \rho v_s \cos \psi \sin \psi \biggl[\frac{\rho^y_s}{\rho^y_n} - \frac{\rho^x_s}{\rho^x_n}\biggr] 
\end{align}
in the definition~\eqref{eq:jE} of the energy current. Here, we abbreviate the $xx$ and the $yy$ components of the superfluid 
and normal tensor by a superscript $x$ and $y$, respectively. Collecting all terms yields
\begin{align}
- \frac{\omega}{q} \delta \varepsilon + \hat{\bf q} \cdot {\bf j}^\varepsilon &= - \frac{\omega}{q} \delta q - \tilde{s} T \biggl\{
g_T \cos \psi \sin \psi \biggl[\frac{\rho^x_s}{\rho^x_n} - \frac{\rho^y_s}{\rho^y_n}\biggr]
\nonumber \\
&\qquad- g_L \biggl[\frac{\rho^y_s}{\rho^y_n} \cos^2 \psi + \frac{\rho^x_s}{\rho^x_n} \sin^2 \psi\biggr] + \rho v_s \biggl[\frac{\rho^y_s}{\rho_{n}^{y}} \cos^2 \psi + \frac{\rho^x_s}{\rho^x_n} \sin^2 \psi\biggr] 
\biggr\} = 0 .
\end{align}
In a similar way, the equation for the gradient of the layer phase becomes
\begin{align}
- \frac{\omega}{q} u + v_{n,y} &= - \frac{\omega}{q} u + \frac{\cos \psi}{\rho} (g_T - j_{s,T}) + \frac{\sin \psi}{\rho} (g_L - j_{s,L}) \nonumber \\
&= - \frac{\omega}{q} u + g_L \frac{\cos \psi}{\rho^y_n} + g_T \frac{\sin \psi}{\rho^y_n} - v_s \frac{\rho^y_s}{\rho^y_n} \cos \psi = 0 . \label{eq:appCu}
\end{align}
To simplify the equation~\eqref{eq:C5} for the superfluid velocity, use the differential~\eqref{eq:defdp} to leading order in the velocities 
in order to replace $d\mu$:
\begin{align}
- \frac{\omega}{q} v_s + \frac{1}{\rho} \frac{\partial p}{\partial \rho} \biggr|_{\tilde{s}, v_s, u'} \delta \rho - \frac{\tilde{s}}{\rho c_V} \delta q &= 0 . \label{eq:appCvs}
\end{align}
Collecting the equations~\eqref{eq:density1},~\eqref{eq:appCjL},~\eqref{eq:appCjT},~\eqref{eq:appCu}, and~\eqref{eq:appCvs} gives 
the characteristic equation (abbreviating $c = \cos \psi$ and $s = \sin \psi$)
\begin{align}
\begin{pmatrix}
- \frac{\omega}{q} & 1 & 0 & 0 & 0 & 0 \\
\frac{K}{\rho} & - \frac{\omega}{q}  & 0 & 0 & 0 & - B c^3 \\
0 & 0 & - \frac{\omega}{q}  & 0 & 0 & - B s c^2 \\
0 &
\tilde{s} T \bigl[\frac{\rho^y_s}{\rho^y_n} c^2 + \frac{\rho^x_s}{\rho^x_n} s^2\bigr]
& 
\tilde{s} T s c \bigl[\frac{\rho^y_s}{\rho^y_n} - \frac{\rho^x_s}{\rho^x_n}\bigr] & - \frac{\omega}{q}  & 
- \rho \tilde{s} T \bigl[\frac{\rho^y_s}{\rho^y_n} c^2 + \frac{\rho^x_s}{\rho^x_n} s^2 \bigr]  & 0 \\
\frac{K}{\rho^2} & 0 & 0 & - \frac{\tilde{s}}{\rho c_V} & - \frac{\omega}{q}  & 0 \\
0 &
\frac{c}{\rho^y_n} & 
\frac{s}{\rho^y_n} & 0 &  
- \frac{\rho^y_s c}{\rho^y_n} & - \frac{\omega}{q} 
\end{pmatrix}
\begin{pmatrix}
\delta \rho \\
g_L \\
g_T \\
\delta q \\
v_s \\
u'
\end{pmatrix}
&= 0 .  \label{eq:coefficientmatrix}
\end{align}
The associated characteristic determinant is
\begin{align}
&\omega^6 
+ 
\omega^4 q^2 \biggl[- \frac{K}{\rho} - \frac{B}{\rho^y_n} c^2 - \frac{\tilde{s}^2 T}{c_V} \biggl(\frac{\rho^y_s}{\rho^y_n} c^2  + \frac{\rho^x_s}{\rho^x_n} s^2 \biggr)\biggr] 
\nonumber \\
& +
\omega^2 q^4 \biggl[\frac{B K}{\rho^2} c^2 \biggl(\frac{\rho}{\rho^y_n} c^2 + \frac{\rho^y_s}{\rho^y_n} s^2\biggr) 
+
\frac{\tilde{s}^2 T}{\rho c_V} \bigg( K \biggl(\frac{\rho^y_s}{\rho^y_n} c^2  + \frac{\rho^x_s}{\rho^x_n} s^2 \biggr)
 + B c^2 \biggl(\frac{\rho^y_s}{\rho^y_n} c^2  + \frac{\rho}{\rho^y_n} \frac{\rho^x_s}{\rho^x_n} s^2 \biggr)\biggr)\biggr] \nonumber \\
& + q^6 \biggl[- \frac{B K}{\rho \rho^x_n} \frac{\tilde{s}^2 T}{c_V} c^2 s^2  \biggr(\frac{\rho^y_s}{\rho^y_n } c^2 + \frac{\rho^x_s}{\rho^y_n} s^2\biggr)\biggr] 
= 0 .
\end{align}
In general, there are three distinct propagating modes. One of them is a generalized transverse 
sound mode $\omega_t(\mathbf{q})$, which is the Goldstone mode associated with smectic order
found by Martin {\it et al.}~\cite{martin72}. As noted in Sec.~\ref{sec:2a}, its velocity vanishes in the special case of purely 
parallel and perpendicular propagation. Formally, this is because the last term of the characteristic 
polynomial vanishes in this limit, giving rise to only two sound modes plus two diffusive ones.  
For propagation in the longitudinal direction ($c=1$ and $s=0$), the matrix~\eqref{eq:coefficientmatrix}
reduces to the expression~\eqref{eq:matrix} in the main text 
(the transverse current component decouples and may be omitted). 
In the opposite limit $\psi \to \pi/2$ where propagation is parallel to the smectic layers, we find instead
\begin{align}
\omega^4 + \omega^2 q^2 \biggl[- \frac{K}{\rho} - \frac{\tilde{s}^2 T}{c_V} \frac{\rho^x_s}{\rho^x_n}\biggr] + q^4 \biggl[\frac{K}{\rho} \frac{\tilde{s}^2 T}{c_V} \frac{\rho^x_s}{\rho^x_n}\biggr]&= 0 .
\end{align}
Note that there is no dependence on $B$, $\rho^y_s$, and $\rho^y_n$. We obtain the standard first and second sound modes with speed
\begin{align}
c_1^2(\psi=\pi/2) &= \frac{K}{\rho} \\
c_2^2(\psi=\pi/2) &= \frac{\tilde{s}^2 T}{c_V} \frac{\rho^x_s}{\rho^x_n} ,
\end{align}
which is as expected: in the special case of propagation along the smectic layers, the density wave structure does not affect the 
hydrodynamic modes, which are the same as for a homogenous superfluid. Since the normal fraction $\rho^x_n$ vanishes at low
temperature, the second sound mode becomes ill-defined.

Our results for the spectrum of hydrodynamic modes in a superfluid smectic phase are also 
relevant for imbalanced Fermi superfluids in a situation where the spatial modulation of the
order parameter only appears along a single direction. As pointed out by Radzihovsky and Vishwanath~\cite{radzihovsky09}, the symmetry-breaking 
layer and superfluid phase variables also appear in a description of the Larkin-Ovchinnikov phase of an imbalanced Fermi superfluid, where the mismatch 
$q_0=k_{F\uparrow}-k_{F\downarrow}$ of the two Fermi surfaces results in an uni-directional periodic modulation of the 
complex gap parameter $\Delta_{q_0}$. The associated low energy theory derived by these authors and discussed in much 
more detail in Ref.~\cite{radzihovsky11} is of the form
 \begin{equation}
\mathcal{H}_{\rm LO}=\frac{K_1}{2}\left(\nabla^2 u\right)^2 +\frac{B}{2}\Bigl(\partial_{\parallel} u - \frac{1}{2} (\nabla u)^2\Bigr)^2 + 
\frac{\rho_s^i}{2}\left(\nabla_i\theta_{\rm sf}\right)^2
\label{eq:LO-Goldstone}
\end{equation}
where $i=\parallel$ or $i=\perp$ refer to the directions parallel and transverse to the ordering vector. The two 
Goldstone modes associated with the Hamiltonian density~\eqref{eq:LO-Goldstone} are determined by the elastic constants
$K_1$ and $B$ together with the two different superfluid densities $\rho_s^{\perp}$ and $\rho_s^{\parallel}$ and
a finite compressibility $\chi$ in the form
 \begin{align}
\omega_{\rm sf}(\mathbf{q}) &= \sqrt{\bigl( \rho_s^{\perp} q_{\perp}^2 + \rho_s^{\parallel} q_{\parallel}^2\bigr)/\chi} \\ 
\omega_{\rm sm}(\mathbf{q}) &=\sqrt{\bigl( K_1 q_{\perp}^4 + B  q_{\parallel}^2\bigr)/\chi}  \, .
\label{eq:LO-Goldstone-frequencies}
\end{align} 
The first mode is an anisotropic version of the Bogoliubov-Anderson mode of a neutral superfluid while the smectic phonon $\omega_{\rm sm}(\mathbf{q})$
is unique for the uni-directional LO state. It has a linear spectrum determined by the layer compression modulus $B$ for
wave vectors along the direction of ordering but turns into a mode with quadratic dispersion for $\mathbf{q}\!\perp\!\mathbf{q}_0$. 
The predicted mode structure differs from that of the superfluid smectic phase discussed above, and indeed
there are important differences between both phases. First of all, the uni-directional LO state only exists in the super\-fluid
regime of the imbalanced Fermi gas. The elastic constants $B$ and $K_1$ therefore derive from a single complex order parameter $\Delta_{q_0}$. 
Moreover, in contrast to the superfluid smectic phase of dipolar BECs, it is assumed that the spatial structure in 
$\Delta_{q_0}$ is not associated with a real density modulation and also that the fermionic superfluid has no underlying zero-momentum condensate. 
These assumptions are valid for essentially incompressible Fermi systems in the BCS limit, where the condensate fraction
is exponentially small. They imply that there is no coupling of the symmetry breaking variables $\nabla u$ and $\mathbf{v}_s$ to the particle and
momentum density. The energy density~\eqref{eq:LO-Goldstone} associated with $\nabla u$ and $\mathbf{v}_s$
thus fully determines the spectrum of Goldstone modes\footnote{It is interesting to note that the number of Goldstone modes is reduced to only one in 
a Fulde-Ferrell phase with uni-directional order. In this case, time-reversal symmetry is broken
and there is only a single contribution $(\rho_s/2)\left(  \hat{\mathbf{q}}_0\cdot\mathbf{v}_s -(\hbar q_0/m)\partial_{\parallel} u  \right)^2$ 
to the free energy density rather than two independent ones associated with superfluid flow and elastic distortions. The Fulde-Ferrell state therefore 
realizes a broken relative gauge symmetry as in $^3$He A~\cite{leggett75}.}.
By contrast, the equations of motion~\eqref{eq:coefficientmatrix} that determine the hydrodynamic modes of a superfluid 
smectic phase depend crucially on the coupling between density fluctuations $\delta\rho$ and both symmetry-breaking variables.
The fact that the superfluid order parameter has a negligible coupling to the particle density in the BCS-limit is well known. 
Specifically, it has been shown by Leggett~\cite{leggett65} that even 
in the presence of strong Fermi liquid corrections in the normal state, the compressibility of a neutral Fermi liquid is unchanged by
a transition to superfluidity. More generally, it may be shown~\cite{zwerger20} that the change $\delta n$ in particle density
at the superfluid transition is related to the finite order parameter $|\psi|^2$ by a linear relation $\delta n=\alpha\tilde{\kappa}|\psi|^2$ 
to lowest order. Here, $\tilde{\kappa}=\partial n/\partial\mu$ is the compressibility and $\alpha$ is the coupling constant between density and the order parameter, which is generically of the form
$-\alpha\delta n|\psi|^2$~\cite{son20}. For weak-coupling BECs, 
one has $\alpha\tilde{\kappa}\equiv 1$.  By contrast, for Fermi superfluids, one finds that $\alpha\tilde{\kappa}\simeq (T_c/T_F)^4$
is exponentially small in the BCS limit and even for a unitary gas the dimensionless coupling constant $\alpha\tilde{\kappa}$ turns out to be 
around $0.05$ only~\cite{zwerger20}.     

\section*{References}

\bibliography{bib_smectic}

\end{document}